# SGR DSPH: A BRIDGE BETWEEN DWARF GALAXIES AND DLAs?

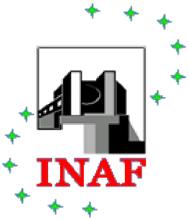
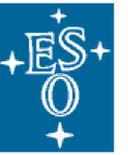

L. Sbordone (ESO – Univ. Roma Tor Vergata), P. Bonifacio (OAT),
G. Marconi (ESO – OAR), L. Pasquini (ESO), V. Hill (Obs. Paris)

## ABSTRACT

We present abundances for 12 giants in the Sagittarius Dwarf Spheroidal Galaxy (Sgr dSph) obtained from VLT-UVES spectra. Moving on a short period, polar orbit around the Milky Way, the Sgr dSph is undergoing tidal disruption and will eventually dissolve in the halo. Our sample is dominated by a metal-rich, $\alpha$-element-poor population, indicative of a long chemical processing of the dSph gas during a slow, probably bursting star formation history. This population is the most metal-rich ever observed in a dwarf galaxy of the local group, and has the lowest $\alpha$-element content. Placing the known abundances of the LG dwarfs on the [$\alpha$/Fe] vs [Fe/H] plane allows now to recognize a well defined evolutive sequence, different from the one followed by the MW disc star, but apparently superimposed to the one followed by many of the Damped Lyman Alpha systems.

## CMD AND SAMPLE SELECTION

Sgr dSph is presently seen through the bulge. As a consequence, its CMD (fig. 1) is highly contaminated, but shows many interesting features: the width of the "main" RGB (violet in fig. 1) is excessive to be explained by photometric errors. A spread in age or in metallicity has to be inferred. A second "RGB-like" structure (red) can be exists above the previous, although there's no confirmation of its membership to Sgr. The so-called "blue plume" (blue) extends above the Turn-Off like a blue straggler population, but is too populated. All these are hints of a complex star formation history, probably related to the strong interaction with the Milky Way. To break the age-metallicity degeneracy in the RGB we have observed a sample of 12 red giants of confirmed membership just above the clump (yellow, and data in table 1),

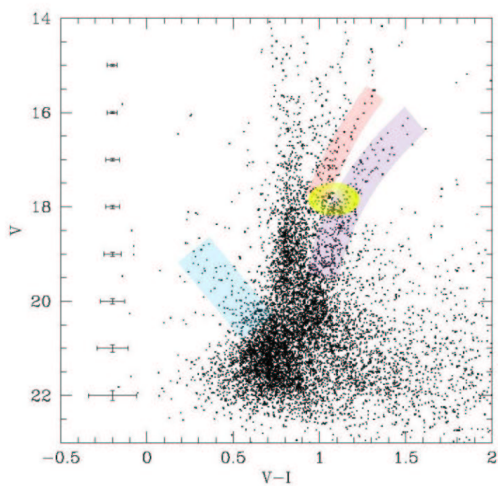

Figure 1: CMD for the "Field 1" of Marconi et Al (1998) in the Sgr dSph (see text for the explanation of the colors)

## DATA REDUCTION AND ANALYSIS

The stars have been observed in dichroic mode D1, (red arm 4800 – 6800 Å) With the exception of stars 772 and 879 (see Bonifacio et al. 2000) we have used pipeline reduced and coadded spectra with a S/N between 19 and 34 around 5300 Å. We have measured EWs for MgI, SiI, CaI, FeI and FeII lines (only red arm have been employed so far). *Ad hoc* model atmospheres have been computed with ATLAS9 (Kurucz 1993) for each star, deriving $T_{eff}$ from the calibration of Alonso et al. (1999) and setting log g from the iron ionization equilibrium. Abundances have been derived from the EWs using WIDTH (Kurucz 1993), and synthetic spectra calculated with SYNTHE (Kurucz 1993) employed to verify the abundances derived from EWs and to derive the oxygen abundance from the weak and blended [OI] 630.0nm line (see fig 2a and b).

## DERIVED ABUNDANCES

The derived abundances are listed in table 2. The entire population we sampled has [Fe/H] > -0.9, while 11 stars are above -0.6. The mean value is -0.3. This is significantly higher than the photometric estimates, giving a [Fe/H] between -1.5 and -0.7.

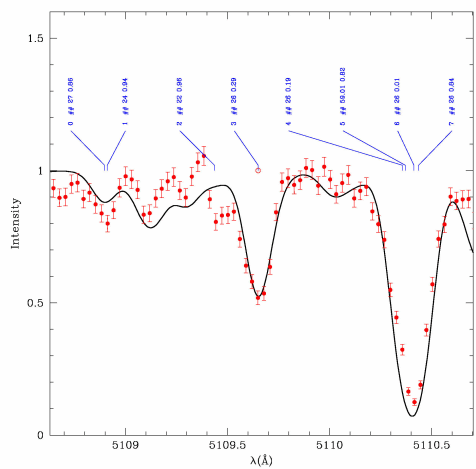

Figure 2: Spectral synthesis for the Fe I 5109.650 Å line (marked by a red open dot) in the star 867, with superimposed the observed spectrum (red filled dots with S/N error bars).

| Star | $\alpha$(2000) | $\delta$(2000) | V | $(V-I)_0$ | $T_{eff}$ | log g | $\xi$ |
|---|---|---|---|---|---|---|---|
| 432 | 18 53 50.75 | –30 27 27.3 | 17.55 | 0.965 | 4818 | 2.30 | 1.3 |
| 628 | 18 53 47.91 | –30 26 14.5 | 18.00 | 0.928 | 4904 | 2.50 | 2.0 |
| 635 | 18 53 51.05 | –30 26 48.3 | 18.01 | 0.954 | 4843 | 2.50 | 1.8 |
| 656 | 18 53 45.71 | –30 25 57.3 | 18.04 | 0.882 | 5017 | 2.50 | 1.6 |
| 709 | 18 53 38.73 | –30 29 28.5 | 18.09 | 0.917 | 4930 | 2.50 | 1.5 |
| 716 | 18 53 52.97 | –30 27 12.8 | 18.10 | 0.902 | 4967 | 2.50 | 2.0 |
| 717 | 18 53 48.05 | –30 29 38.1 | 18.10 | 0.872 | 5042 | 2.50 | 1.3 |
| 772 | 18 53 48.13 | –30 32 0.8 | 18.15 | 0.947 | 4891 | 2.50 | 1.5 |
| 867 | 18 53 53.02 | –30 27 29.2 | 18.30 | 0.933 | 4892 | 2.50 | 2.0 |
| 879 | 18 53 48.59 | –30 30 48.7 | 18.33 | 0.965 | 4891 | 2.50 | 1.4 |
| 894 | 18 53 36.84 | –30 29 54.1 | 18.34 | 0.940 | 4876 | 2.50 | 1.4 |
| 927 | 18 53 51.69 | –30 26 50.7 | 18.39 | 0.937 | 4880 | 2.75 | 1.2 |

Table 1: photometric data and atmospheric parameters for the 12 stars of our sample (from Bonifacio et al. 2003).

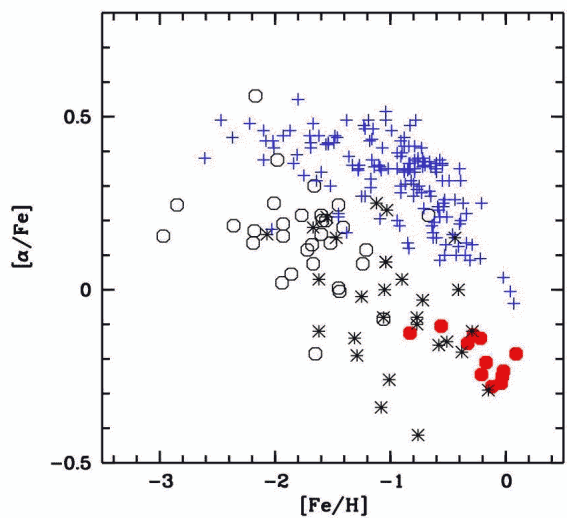

Figure 3: [$\alpha$/Fe] vs. [Fe/H] for the Sgr dSph (red filled dots), others local group dwarf (Shetrone et al. 2001, 2003, open dots), MW disc stars from Gratton et al. (2003) (blue crosses) and DLAs from Centuriòn et al. (2003) (asterisks). From Bonifacio et al. (2003).

| Star | [Fe/H] | [O/FeII] | [Mg/Fe] | [Si/Fe] | [Ca/Fe] |
|---|---|---|---|---|---|
| 432 | –0.83 ± 0.12 | +0.01 | –0.01 | –0.06 | –0.24 |
| 628 | –0.22 ± 0.11 | –0.16 | –0.06 | –0.02 | –0.22 |
| 635 | –0.33 ± 0.12 | –0.02 | –0.05 | –0.14 | –0.26 |
| 656 | –0.17 ± 0.10 | –0.18 | –0.24 | –0.09 | –0.18 |
| 709 | –0.02 ± 0.20 | –0.14 | –0.25 | –0.21 | –0.22 |
| 716 | –0.12 ± 0.11 | –0.16 | –0.24 | –0.17 | –0.32 |
| 717 | +0.09 ± 0.11 | –0.09 | –0.16 | –0.15 | –0.21 |
| 772 | –0.21 ± 0.19 | –0.11 | –0.23 | –0.07 | –0.26 |
| 867 | –0.56 ± 0.19 | –0.01 | –0.13 | +0.01 | –0.08 |
| 879 | –0.28 ± 0.16 | ≤ 0.18 | –0.05 | –0.07 | –0.21 |
| 894 | –0.04 ± 0.14 | +0.11 | –0.34 | +0.00 | –0.20 |
| 927 | –0.03 ± 0.15 | –0.09 | –0.29 | –0.10 | –0.20 |

Table 2: Derived abundanced and uncertainties for the 12 Sgr dSph giants. From Bonifacio et al. (2003)

More striking, the $\alpha$ to iron ratio appear to be generally subsolar, even at the lowest metallicities, with strong hints of a trend towards lower $\alpha$ content at higher metallicities.

## INTERPRETATION

Sgr dSph appears to host a metal rich population the composition of which appears to be characteristic of a gas enriched during a long and slow (or bursting) star formation history. In our view, the most likely explanation is provided by a sequence of starbursts triggered by the disc crossings experienced by the dSph once every Gyr. In fact, by using the derived metallicities and the Isochrones by Girardi et al. (2002), we infer a very young age (less than 2 Gyr) for the observed stars. This is coherent with the timescale of he last disk crossings, and lead to interpret the "blue plume" as the ZAMS of the same population.

## A LINK WITH THE DLAs ?

In fig. 3 we show in a [$\alpha$/Fe] vs. [Fe/H] diagram our Sgr data together with measures for other local group dwarfs from Shetrone et al. (2001, 2003), galactic disk stars from Gratton et al. (2003) and DLAs from Centuriòn et al. (2003). For stellar data, $\alpha$ is defined as the mean of Mg and Ca, while for the DLAs the [Si/Zn] ratio is used instead of [$\alpha$/Fe]. As can be clearly noted, dwarf galaxies appear to follow a common "chemical path" in this scheme, with the Sgr dSph population here analyzed filling the iron-rich – $\alpha$-element-poor end of the sequence. Moreover, the stars of the MW's disc appear to occupy a well separated locus in the diagram. It is also intringuing to note how the majority of the DLAs observed by Centuriòn et al. (2003) appear to lie on the same sequence followed by the LG dwarf galaxies. It is also tempting to conclude that dwarf galaxies share a common evolutionary path with most of the DLAs, or, even more, that most of the DLAs are in fact dwarf galaxies observed when still gas-rich. A slow star formation rate in DLAs, resembling the one typical of dwarf galaxies, has been put forward by Centuriòn (2000). Figure 3 provides a strong evidence supporting this link.